\documentclass[fleqn,usenatbib]{mnras}
\usepackage{newtxtext}

\usepackage[T1]{fontenc}
\usepackage{ae,aecompl}
\usepackage{graphicx}   % Including figure files
\usepackage{amsmath}    % Advanced maths commands
\usepackage{amssymb}    % Extra maths symbols
\newcommand{\hinv}{h^{-1}}
\newcommand{\mpc}{\rm{Mpc}}
\newcommand{\hmpc}{\hinv\mpc}
\newcommand{\kpc}{\rm{kpc}}
\newcommand{\hkpc}{\hinv\kpc}

\newcommand{\hmsol}{\hbox{$\hinv {\rm M}_\odot$}}

\title[Spin-Filament relation]{Relation between halo spin and cosmic web filaments at $z \simeq 3$}

\author[Gonz\'alez et al.]{
Roberto E. Gonz\'alez,$^{1,2}$\thanks{E-mail: regonzar@astro.puc.cl}
Joaquin Prieto,$^{3}$
Nelson Padilla$^{1,2}$
and Raul Jimenez$^{4,5}$
\\
$^{1}$Instituto de Astrof\'{i}sica, Pontificia Universidad Cat\'olica, Av. Vicu\~na Mackenna 4860, Santiago, Chile\\
$^{2}$Centro de Astro-Ingenier\'{i}a, Pontificia Universidad Cat\'olica, Av. Vicu\~na Mackenna 4860, Santiago, Chile\\
$^{3}$Departamento de Astronom\'{i}a, Universidad de Chile, Casilla 36-D, Santiago, Chile \\
$^{4}$ICREA \& ICC, University of Barcelona, Marti i Franques 1, Barcelona 08028, Spain \\
$^{5}$Radcliffe Institute for Advanced Study, Harvard University, MA 02138, USA
}

\date{Accepted XXX. Received YYY; in original form ZZZ}

\pubyear{2016}

\begin{document}
\label{firstpage}
\pagerange{\pageref{firstpage}--\pageref{lastpage}}
\maketitle

% Abstract of the paper 250 words max
\begin{abstract}
We investigate the spin evolution of dark matter haloes and their dependence on the number of connected filaments from the cosmic web at high redshift (spin-filament relation hereafter). To this purpose, we have simulated $5000$ haloes in the mass range $5\times10^{9}$\hmsol$\,$ to $5\times10^{11}$\hmsol$\,$ at $z=3$ in cosmological N-body simulations. We confirm the relation found by \citet{2015MNRAS.452..784P} where haloes with fewer filaments have larger spin. We also found that this relation is more significant for higher halo masses, and for haloes with a passive (no major mergers) assembly history. Another finding is that haloes with larger spin or with fewer filaments have their filaments more perpendicularly aligned with the spin vector. Our results point to a picture in which the initial spin of haloes is well described by tidal torque theory and then gets subsequently modified in a predictable way because of the topology of the cosmic web, which in turn is given by the currently favoured LCDM model.
Our spin-filament relation is a prediction from LCDM that could be tested with observations.
\end{abstract}

% Select between one and six entries from the list of approved keywords.
% Don't make up new ones.

%%%%%%%%%%%%%%%%%%%%%%%%%%%%%%%%%%%%%%%%%%%%%%%%%%
\section{Introduction}

In the current cosmological paradigm-the Lambda Cold dark Matter (LCDM) model- the angular momentum of dark matter (DM) haloes is
originated due to
the misalignment between their moment of inertia and the tidal tensor
associated to the
mass distribution around them. The coupling between these two terms is
responsible for the
original spin of DM haloes.

In his seminal paper \citet{1951pca..conf..195H} realized that galaxies could acquire
their spin
via tidal torques associated to neighbor galaxies. Later, \citet{1969ApJ...155..393P}
and \citet{1970Ap......6..320D}, placed Hoyle's idea in the hierarchical galaxy formation framework. This effort ended in 
 the so called tidal
torque theory, TTT \citep{1969ApJ...155..393P,1970Ap......6..320D,1984ApJ...286...38W,1988MNRAS.232..339H}. However, 
it is still unclear what sets the halo final spin (at $z \sim 0$) after the merger and baryonic processes take place.

Motivated by the work of \citet{1999AA...343..663P} recent interest has
been focused on the vorticity generation
in the large scale structures of our Universe. The authors showed that shell crossing associated
to the collapse of DM filaments at large ($\sim$ few Mpc) scales generates a quadrupole vorticity
pattern parallel to the filaments. This effect has been later seen in 
 larger scales simulations \citep{2015MNRAS.446.2744L}.
In this scenario, small DM haloes inside any vorticity quadrant should have a spin%%%NELSON: no entiendo lo de las celdas
parallel to its host filament, whereas bigger
DM haloes covering more than one vorticity quadrant should tend to have a spin
perpendicular to its host filament.

In line with the above studies, using a cosmological simulation
\citet{2012MNRAS.427.3320C} have shown that
DM haloes have an intrinsic alignment (IA) with the cosmic web: small
haloes tend to be parallel to their host DM
filament and big haloes tend to be perpendicular to their filament. They
found a characteristic redshift dependent
mass for the alignment transition from a perpendicular to a parallel
orientation with respect to the host filament.
Such scale mass has a direct connection with the quadrupole vorticity scale
size predicted by \citet{1999AA...343..663P}.
\citet{2015MNRAS.446.2744L} have extended
the IA studies to the stellar component of
their cosmological simulations finding consistent results: low mass
galaxies are aligned with the filaments and
high mass galaxies tend to be perpendicular to the filaments \citep{2007MNRAS.375..489H,2013ApJ...766L..15L,2014MNRAS.444.1453D}.

{

Other studies also show spin and galaxy alignments with filaments but at large scales $>1\hmpc$ \citep{2015MNRAS.454.2736C, 2015MNRAS.448.3522T, 2015MNRAS.454.3328V, 2015MNRAS.452.3369C}. Our study is focused on smaller scales, in particular the halo-filament connectiviy at few virial radii.

}

\citet{2007MNRAS.375..489H} found that low-mass haloes in clusters have a higher median spin than in
filaments and present a more prominent fraction of rapidly spinning objects.  The environment of these objects suggests that the reason for the enhanced spin is likely due to recent major mergers.

More recently, \citet{2015arXiv151001711B} studied the relationship between changes in the orientation of the angular momentum vector of dark matter haloes ("spin flips") and found spin flips could be a mechanism for galaxy morphological transformation without involving major mergers.  One third of haloes have, at some point in their lifetimes, experienced a spin flip of at least $45\deg$ that does not coincide with a major merger.

On the observational side, \citet{2002ApJ...567L.111L} studied alignments in the IRAS
Point Source Catalogue Redshift survey (PSCz), and found that the galaxy spin is typically aligned perpendicular to
filaments. \citet{2010MNRAS.404..975J} showed how star formation in galaxies and spin are correlated, with a $5\sigma$ detection of spin alignment among galaxies which formed most of their stars at $z > 2$. \citet{2016MNRAS.457..695P} also found an alignment of galaxy spin with the shear field in the 2MASS Redshift survey. On the other hand, \citet{2013ApJ...775L..42T} observed a parallel alignment between the spins of bright spiral
galaxies and filaments, while early-type (mostly lenticular) galaxies had their spins aligned perpendicular to the filament direction. 
The halo spin and comic-web alignment may depend on halo mass, cosmic-web definitions, environment and spin flips.
A thorough review on this topic can be found in  \citet{2015SSRv..193....1J}.

Using high resolution cosmological hydro-dynamical simulations
\citet{2015MNRAS.452..784P} studied in detail the spin acquisition process
of four $10^9M_\odot$ DM haloes with very different spin parameters at
redshift $z=9$. They show that low spin haloes were located
in knots of the cosmic web whereas the high spin haloes belonged to
filaments inside walls. Such a spin-environment relation motivated them to
run DM only simulations; they found that the spin of DM haloes anti-correlates with the
number of local filaments converging onto the
halo for a redshift $z=9$: the greater the number of filaments the lower
the spin parameter. Such a finding strengthens the fact that DM haloes are
not independent structures inside the cosmic web but rather their properties should
be related to their surrounding structures at scales of $\sim$ few $R_{vir}$.
In \citet{2015MNRAS.452..784P} the spin-number of filaments
anti-correlation was understood as the result of isotropic
angular momentum cancellation, i.e. for a large number of converging
filaments the net spin is more likely to cancel when coming from different 
directions \citep{2002ApJ...581..799V}.

In order to investigate further the spin-number of filaments relation, we have produced new
DM simulations covering a wider mass range
and reaching lower final redshifts than \citet{2015MNRAS.452..784P}. This has allowed us to further confirm the anti-correlation between spin and number of filaments, but also to discover new effects due to mass and alignment with the cosmic web filament.
%The paper is organized as follow:

\section{Data}
\subsection{Simulations}

We use two simulations to cover two orders of magnitude in mass at redshift $3$.
The simulations were run using the Gadget2 code \citep{2005MNRAS.364.1105S} with cosmological parameters from Plank13 \citep{2014A&A...571A..16P}, where $\Omega_0=0.3175$, $\Omega_{\Lambda}=0.6825$, $\Omega_b=0.049$, $h=0.6711$. Both simulations consist of $1024^3$ particles, but one has a box size of $20\hmpc$ and the other $40\hmpc$ with particle masses $6.56 \times 10^5$\hmsol$\,$ and $5.25 \times 10^6$\hmsol$\,$ respectively.

In the smaller simulation we can resolve $\sim2500$ haloes at $z=3$ in the mass range 
$5\times10^{9}$\hmsol$\,$ to $5\times10^{11}$\hmsol$\,$ with at least $\sim 10000$ particles within their virial radius.
For the larger simulation we can resolve a similar number of haloes but eight times more massive with the same number of particles.
This allows us to compute reliable spin values, and more important, to resolve the filamentary structure around haloes at $\sim$ few $R_{vir}$.

Halos and merger trees are obtained using the Rockstar Halo Finder and Consistent Tree codes \citep{2013ApJ...762..109B}.

\subsection{Filaments}

For each halo, we extract particles within $3$ virial radii and run the \small{DISPERSE} code \citep{disperse} on the particle distribution. 
We choose \small{DISPERSE} because it successfully recovers filamentary structures even on coarse particle distributions; it uses a simple definition for filaments based on a single parameter, the persistence threshold, and it has been used in several simulations and observational data sets showing that resulting filaments are well correlated with filamentary structure \citep{2012MNRAS.427.3320C,2015MNRAS.446.2744L,2016A&A...590A.110C,2016arXiv160507434A,2016arXiv160106434G}.

{ 
The acronym \small{DISPERSE} stands for Discrete Persistent Structures Extractor.  The code performs the identification of persistent topological features such as peaks, voids, walls and in particular filaments which are of interest for our study.
\small{DISPERSE} uses Morse Theory, where structures are identified as components of the Morse-Smale complex of an input function defined over a given manifold. 
The input function, usually referred to as Morse function, captures the relationship between the gradient of the function, its topology and the topology of the manifold it is defined over.

In Morse Theory there are critical points and integral lines. Critical points are located where the gradient of the function is null; in our case the function is the density, and the critical points can be: minima (index=0), saddles (index=1) and maxima(index=2) points. 
Integral lines are curves tangent to the gradient field at every point.

Integral lines produce a tessellation of the space into regions called ascending (or descending) k-manifolds where all the field lines originate (or lead).
A k-manifold tessellation produces regions of k dimensions which depend directly on the critical point index used. 

Voids are obtained computing the ascending 3-manifolds, walls by ascending 2-manifolds, and filamentary skeletons by ascending 1-manifolds, which are just the set of arcs connecting two maximum points by several saddle points, it is also called the upper skeleton.

Persistence is a measure of the robustness of the topological features, indicating the level of contrast of features with respect to their background.
Persistence is defined as the ratio between the value (density in this case) of two critical points with different critical indices.  Values of persistence 
close to the noise indicate sensibility to changes in the value of the morse function, and the corresponding critical points can be easily destroyed even by small perturbations.  
}

We run \small{DISPERSE} with a persistence threshold of $8\sigma$ to ensure that only strong filaments are used.
The persistence threshold defines the strength of filament arcs, and this is measured as the number of sigmas above a random distribution. A higher or lower threshold does not change the trends found in this paper neither our conclusions, the only change is that using a threshold below $\sim 6\sigma$ adds too many spurious filaments, with the effect of increasing the average number of filaments connected to each halo changing the normalization of the spin and number of filament relation, but not the trend or slope.
A more detailed description of this detection threshold and resolution/robustness tests can be found in \citet{2016arXiv160106434G}.

The definition of a filament is still not a settled matter in cosmology; there are several definitions and methods for filament detection, and discussing 
which method/definition is better is beyond the scope of this paper.
We are aware that using different persistence threshold filament definition may lead to different number of filaments connected to each host. Therefore, a possible fair comparison with other filament detection methods must, as a minimum, consider only the N strongest filaments so that the average number of filaments connected to each host is the same as in this present study. We use a high detection threshold to use only the strongest filaments and stay in the safe zone of convergence among several methods and visual inspection \citep{2012MNRAS.427.3320C,2015MNRAS.446.2744L,2016arXiv160106434G}.

{
After finding the filaments, we keep filaments that satisfy (i) being closer than $0.5R_{VIR}$; (ii) that it extends beyond the virial radius; (iii) its disconnected arcs or branches are merged into the main filament if they are closer than and angular separation of $15$ degrees from the main filament, where the angle is measured from the halo center. In this way we discard any tangential filament not connected to the halo, and ensure filaments coming from the outskirts and reaching al the way into the halo center.  We also merge close filaments/arcs/branches to avoid repeated counts. 
}

{ In addition, we rank the filament by their
 particle density to obtain the strongest/prominent filament connected to each halo. This density is computed by taking a cylinder in the average direction of the filament, starting from $0.1$ to $3$ virial radius and with a cylinder radius of $0.1R_{VIR}$.}

\begin{figure}
    \includegraphics[width=0.98\columnwidth,natwidth=1024,natheight=953]{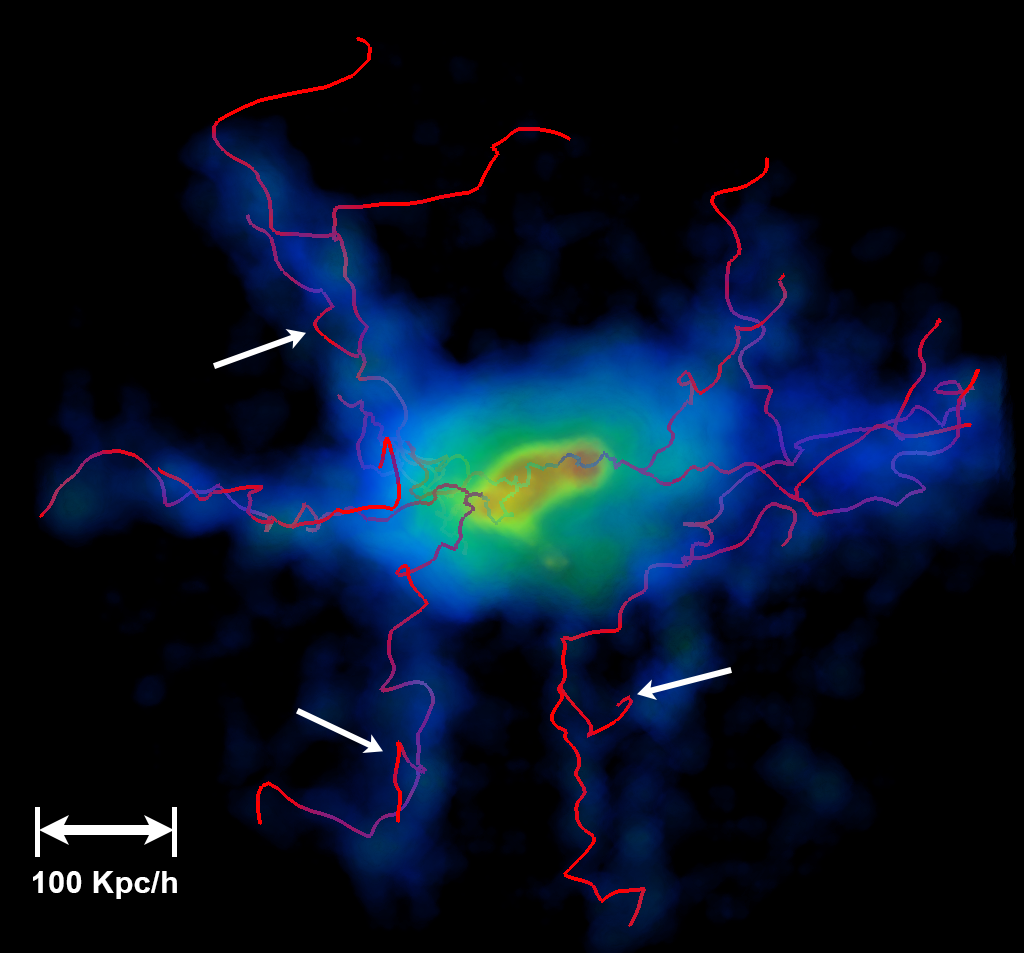} %
    \caption{ DM density projection for one of the host haloes within 3 virial radii.
% using two different color maps. 
\small{DISPERSE} filaments are shown in red
% and blue(right) lines, 
and they trace the DM filaments around haloes.
{ Size scale is indicated in the key; white arrows indicate examples of branches comming out from main filaments.}
}
\label{figfils}
\end{figure}

In figure \ref{figfils} we show the DM density projection around one host {halo of mass $2.6\times10^{11}$\hmsol$\,$} out to $3$ virial radii, using a 3D volume rendering technique; \small{DISPERSE} filaments are shown in colored lines. { The figure shows a box of $800\hkpc$ a side. 
In this case, the connectivity criterion results in $6$ filaments, where several branches (indicated by white arrows) are merged into the main filament; a more difficul scenario happens in the right side of the figure were several branches appear to be interconnected, but our criterion selects two main filaments.}
%Both images show the same with different color maps to make it clear by eye that detected filaments follow higher DM density paths.

\section{Results}
\subsection{Filament/Host properties}

\begin{figure}
    \includegraphics[width=\columnwidth]{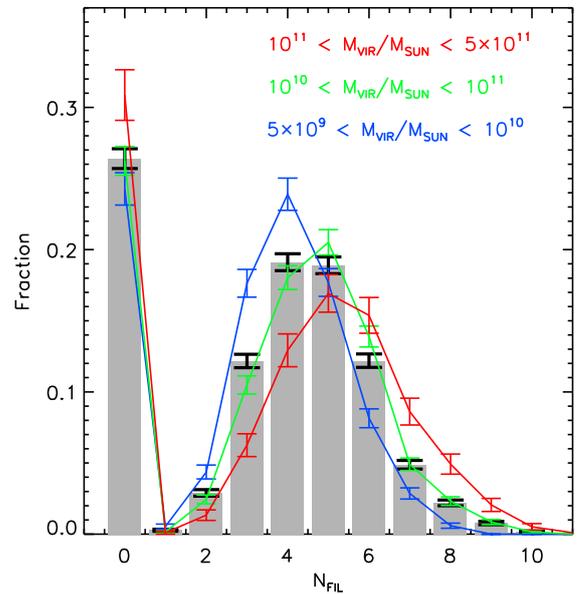} 
    \caption{ Distribution of number of filaments for all host haloes { (gray histogram)}.  Colors correspond to subsamples of different host masses.
}
\label{fignfil}
\end{figure}

We first explore the frequency of filaments around our host haloes.
In figure \ref{fignfil} we show the distribution of the number of filaments for all DM haloes. We find  that $\sim25\%$ of DM haloes have no filaments. There are almost zero haloes with one filament; this is expected since it is very unlikely to form a node at the edge of a void; they do form in the intersection of walls an filaments. The average number of filaments connected to each halo is $\sim 4$.
The more massive the haloes, the more filaments we find are connected to them; this is expected since more massive haloes are more likely to be located at the intersection of filaments and walls.

\begin{figure}
    \includegraphics[width=\columnwidth]{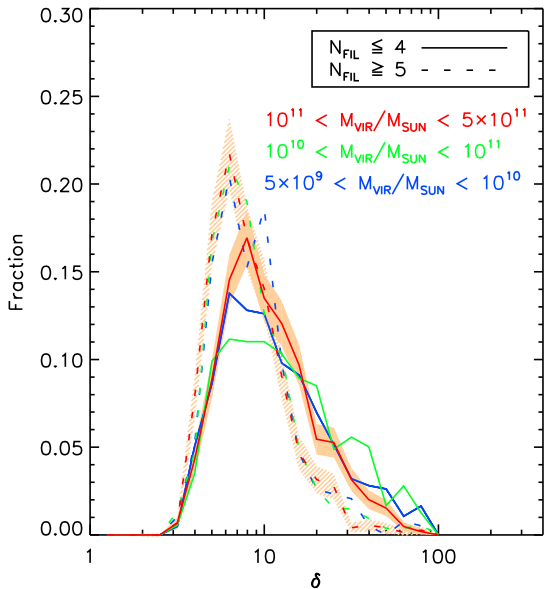} 
    \caption{ DM outskirt overdensity distribution for host haloes with different number of filaments. Colors show distributions for different masses.
{ Solid lines indicate halos with $4$ or less filaments, and dashed lines indicate halos with $5$ or more filaments. Colored regions show errors for the higher mass bins.}
The overdensity is computed for each halo within a shell of $1$ to $3$ virial radii. Halos with fewer filaments are associated to larger outskirt overdensity, probably due to more diffuse accretion occurring in lower local densities, and asymmetric filamentary accretion in larger local densities.
}
\label{figdens}
\end{figure}

We compute the local overdensity around haloes within a shell of $1$ to $3$ virial radii. The typical overdensity in such regions is $\sim 10$.
Figure \ref{figdens} shows overdensity distributions for haloes with few ($\leq 4$) and several ($\geq 5$) filaments{ (solid and dashed lines respectively)}, and for different DM halo masses. We found that haloes with fewer filaments are associated to larger overdensities and this behavior is independent of halo mass.
This result may be in contradiction with the scenario where more filaments implies higher density, however this can be explained due to the fact that haloes with few filaments have on average stronger filaments with higher density contrast. Therefore, more diffuse accretion (several filaments) is associated to lower local densities and asymmetric filamentary accretion (few filaments) with larger local densities.

\subsection{Number of Filaments and Spin}

We have explored the spin-filament relation previously found in \citet{2015MNRAS.452..784P}, but using a larger sample, a wider redshift range and a larger range of masses. 
The halo spin definition used in this paper is the same as in  \citet{2001ApJ...555..240B}:

\begin{equation}
\lambda = \frac{J}{\sqrt{2}MVR}
\end{equation}

\noindent where $J$ is the angular momentum inside a sphere of radius $R$ containing mass $M$, and where $V$ is the halo circular velocity at radius $R$, $V^2=GM/R$.

\begin{figure}
    \includegraphics[width=\columnwidth]{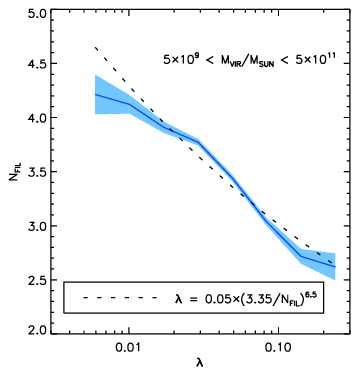}
    \caption{Spin-filament relation at $z=3$. Results combine $\sim 5000$ haloes from two simulations covering the mass range $5\times10^{9}$\hmsol$ < M_{VIR} < 5\times10^{11}$\hmsol.
The Jackknife error amplitude is shown in blue.
The dashed line shows the power-law fit suggested by \citep{2015MNRAS.452..784P}, although with a different normalization and a slope $\sim6.5$. 
This relation remains present at least in the redshift range $10<z<3$ where major mergers are still not the dominant contributor to spin growth.}
    \label{fig1}
\end{figure}

We confirm that haloes with larger spin have less connected filaments. The scatter for individual haloes is large, however the average for large samples shows a tight correlation.  Figure \ref{fig1} shows the spin-filament relation, errors bands are shaded in blue. The dashed line indicates the power-law fit proposed by \citet{2015MNRAS.452..784P}, but with a different normalization given that here we use a higher detection threshold.

This anti-correlation can be understood in a scenario where halo accretion from fewer preferred filaments/directions may boost the angular momentum increase in comparison with several filaments/directions where angular momentum contributions from different directions cancel each other.
In addition, we expect this relation to be more marked at high redshift or, for recently forming haloes, where major mergers are not important yet, since single major mergers can suddenly change the spin magnitude and direction.

\subsection{Dependence on halo mass}

\begin{figure}
    \includegraphics[width=\columnwidth]{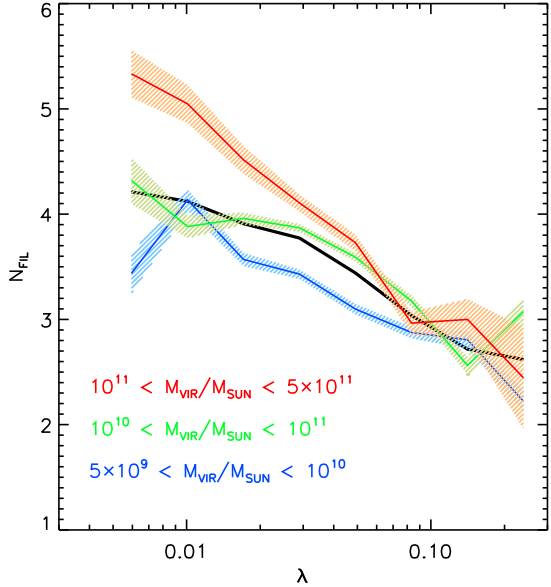}
    \caption{Same as previous figure but for different mass ranges. There is a dependence on mass; more massive haloes have more filaments and the slope of the spin-filament relation is steeper.}
    \label{fig2}
\end{figure}

In Fig.~\ref{fig2} we explore the spin-filament relation for three ranges of mass. For all masses we find the same trend of less filaments for larger spins, however in more massive haloes there are more filaments and a steeper slope. 
We compute the power law fits for these three mass ranges, where $\lambda=0.05\times(a/N_{FIL})^{b}$. 
Resulting parameters from lower to higher halo mass ranges are $a=[3.1,3.5,3.6]$, and $b=[7.0,7.8,4.8]$.
In the higher spin regime, the three spin-filament curves overlap and seem to be independent of halo mass. 
In the low spin case there are larger differences with mass which could be explained in the scenario that low spin and low mass haloes may be very young haloes with too little time to grow their angular momentum even in favorable spin growth conditions; On the other hand, in the same scenario, low spin and high mass haloes have formed earlier, with enough time to grow their spin, but this does not take place because they have too many filaments, or due to mergers.  

{
The figure shows that there are more filaments in more massive halos, and in Figure $3$ it can be seen that halos with less filaments have higher outskirt densities; this may lead to the conclusion that more massive halos are located in less dense environments in apparent contradiction with the standard scenario of structure growth where more massive halos form in denser environments.
However, we clarify that Figure $3$ does not show significant mass dependence on the outskirt density, it only shows that halos with fewer filaments live, on average, in more dense outskirts because those few filaments are thick and dense.  As halos with stronger filaments are distributed among the three mass cuts, a fair comparison to show the mass/density relation should be done at a fixed number of filaments.
}

\subsection{Dependence on halo assembly history}

\begin{figure*}
    \includegraphics[width=0.66\columnwidth]{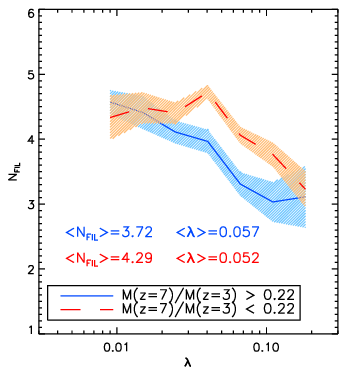}
    \includegraphics[width=0.66\columnwidth]{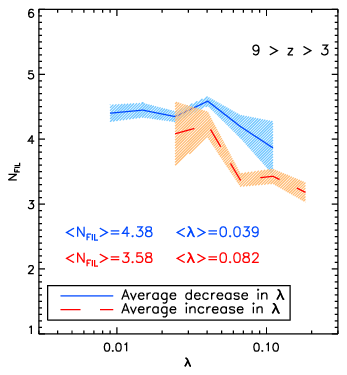}
    \includegraphics[width=0.66\columnwidth]{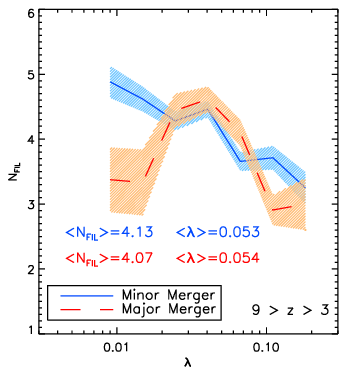}
    \caption{Spin-filament relation for different assembly histories. Left:  Mass ratio between $z=7$ and $z=3$, larger values (blue) correspond to little fractional mass growth within that period; small values (red) indicate large fractional mass growth. Center: Decrease or increase in the spin within the redshift range $9>z>3$. Right: Merger history for $9>z>3$; a merger is considered major if the mass ratio is above $1:5$.
We have $28\%$ of these haloes suffered a major merger. 
Halos with lower mass growth rate, with growing spin, and with no major mergers, show a clear and consistent spin-filament relation, however haloes with strong mass changes, with a decrease in their spin evolution, or haloes suffering major mergers, show a disrupted spin-filament relation. This is consistent with the scenario that major mergers produce large changes in the halo spin, diluting the spin-filament relation.}
    \label{fig3}
\end{figure*}

We study how the assembly properties of the haloes in the redshift range $9>z>3$ affect the spin-filament relation, and show results in Figure \ref{fig3}.
{ The average values of the number of filaments and the spin are shown for each sub-sample.}
In the left panel, we divide the haloes in two samples based on how much their mass increased from $z=7$ until $z=3$; the average of the mass ratio at these two epochs is $0.22$; i.e., on average haloes increased their mass $\sim 5$ times within that period.  Halos with a lower mass increase (blue) show a clear spin-filament relation with the same slope as in Figure \ref{fig1};  haloes with large mass increase(red), probably due to mergers, show a disrupted spin-filament relation, but still with an average negative slope.

In the central panel, we divide into two samples of haloes which on average increase in their spin(red), and decrease their spin (blue).  There is a flat spin-filament relation for haloes with a decreasing spin consistent with a zero slope, a reduction that may be associated again to mergers dissipating angular momentum;  haloes which increase their spin show a spin-filament relation with the same slope as in Figure \ref{fig1}. Notice that in the latter case, we do not plot the curve for lower spin since there are not enough haloes that satisfy at the same time an increase in their spin and a low spin value.
{ Halos that lose spin have a larger number of connected filaments, indicating that diffuse isotropic accretion tends to reduce the spin.}

In the right panel we explore the effect of mergers, where major mergers are defined by a merger ratio of $1:5$.  The sample of haloes affected by major mergers show a disrupted spin-filament relation due to random spin changes produced by mergers, consistent with an average zero slope; on the other hand,  haloes affected only by minor mergers show a strong spin-filament relation with the same slope as in Figure \ref{fig1}.  
{ This is in agreement with \citet{2015MNRAS.452..784P} who show that major mergers may change dramatically the spin magnitude and direction. A major merger may ocurr from any random direction not necessarily from a connected filament; it may also boost or reduce the spin depending on whether the merging halo reaches the host's center in a co- or counter-rotating fashion. Finally, after the merger the host halo is non-relaxed and this may contirbute to the erasing of the spin-filament relation.

In addition, we have that the effect of mergers disrupting the spin-filament relation is stronger for low spin halos ($\lambda<0.04$), and we recover the monotonic general relation at larger spin values ($\lambda>0.04$). This is in agreement with the scenario that for halos with strong filamentary accretion where spin was boosted, mergers also are likely to come from those strong filaments and this may even enhance or maintain the spin-filament relation without disrupting it.}
We can conclude that the scenario of filaments regulating the spin growth of haloes is only valid in the early stages of halo formation before major mergers begin influencing the fate of the halo spin.

\subsection{Dependence on halo redshift}

{ We compute spin-filament relation for the large simulation only, and at three different redshifts $z=[3,4,5]$.
We found there is no significant evolution in the slope of the spin-filament relation. However, there is an increase in the total number of filaments at lower reshifts. This can be explained because the average mass of host halos increases with time.

We expect a decrease in the slope of the spin-filament relation at lower redshift below $z=2$ where major mergers become the main drivers of halo mass evolution. However, to reach lower redshift would require new sets of simulations with higher volume at the same resolution, which is the subject of a forthcoming paper (Gonzalez et al. in prep.).
}

\begin{figure}
    \includegraphics[width=\columnwidth]{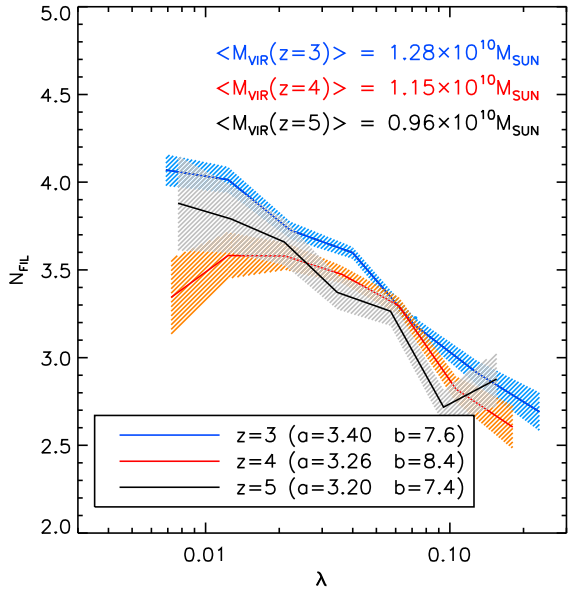}
    \caption{
{  Spin-filament relation at different redshifts. There is no significant change in the slope although a slight increase of the total number of filaments is observed at lower redshifts, produced by the increase of the average host mass.}
}
    \label{figredshift}
\end{figure}

\subsection{Alignment}

\begin{figure}
    \includegraphics[width=\columnwidth]{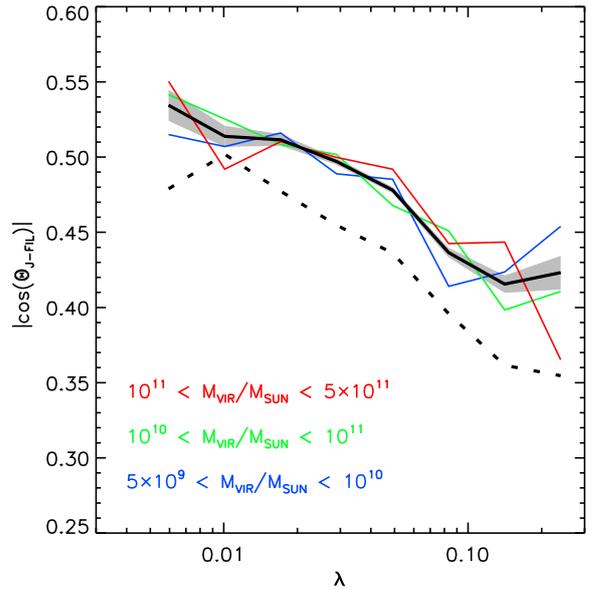}
    \caption{Relation between the spin and the cosine of the average angle between the halo angular momentum and their connected filaments. Results are shown for all masses (solid black) and different masses (see the figure key). 
There is a clear trend where higher spins are associated to filaments more perpendicular to the angular momentum, and there is no noticeable dependence on halo mass.
 This is in agreement with the scenario where filaments perpendicular to the angular momentum contribute more to the angular momentum growth.
In dashed lines we show the relation between the spin and the alignment angle using the most prominent filament;  we find it is in general more perpendicularly aligned to the spin than using all connected filaments.
}
    \label{fig6}
\end{figure}

We now explore the alignment between the halo spin vector and the filament directions.
For this purpose we define the alignment angle $\Theta_{J-Fil}$ as the angle between the halo angular momentum vector and the direction of a connected filament. 
Therefore, we compute the cosine of the alignment angle for each halo using all its connected filaments, and we also compute the cosine of the alignment angle using only the most prominent filament defined in section $2.2$. 

In figure \ref{fig6} we show the relation between the cosine of the alignment angle and the spin using the average of all connected filaments (solid black line), and for the most prominent filament (dashed black line).  
We found that the higher the spin then more perpendicular to the spin their filaments are, and this relation does not depend on mass.
The alignment using the most prominent filament of each halo shows the same trend. However, regardless of spin value, the most prominent filament is more perpendicular to the spin when compared with the average angle using all connected filaments.
This is in agreement with an optimal angular momentum boost from perpendicular filamentary accretion \citep{2015MNRAS.452..784P}, and is consistent with a perpendicular alignment between spin and closest filament shown by \citet{2012MNRAS.427.3320C,2016MNRAS.457..695P} for the most massive haloes.

\begin{figure}
    \includegraphics[width=\columnwidth]{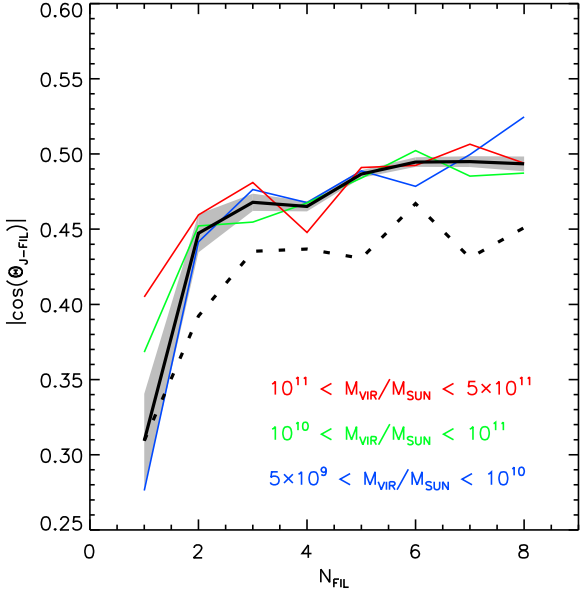}
    \caption{Same as previous figure but using the number of filaments rather than spin in the x-axis. 
Halos with less filaments have a better perpendicular alignment to the spin; for a larger number of filaments this relation weakens. This is in agreement with the scenario where higher spins are associated to  less filaments which are more perpendicularly aligned to the angular momentum.}
    \label{fig7}
\end{figure}

Figure \ref{fig7} shows the relation between the alignment angle and the number of connected filaments. Lines have the same definition as in Figure \ref{fig6}.  Halos with fewer filaments show more perpendicular alignments, and this signature is very strong for single filament haloes. There is no significant dependence on halo mass. These results seem to be in contradiction with \citet{2012MNRAS.427.3320C}
for small mass haloes inside the vorticity quadrants, however we note that our filaments are not large scale $\sim$ Mpc filaments but rather local structures of
$\sim$ few $R_{vir}$ around the DM haloes.

These results are in good agreement with the scenario that strong angular momentum growth comes from accretion from a main filament where the angular momentum aligns perpendicular to the filament, and additional filaments misaligned with the main filament tend to cancel out some of the contributions to the angular momentum.

\section{Discussion}

We explore the spin-filament relation previously found in \citet{2015MNRAS.452..784P}, using a larger sample with a broader range of masses and down to lower redshifts.
We confirm that haloes with larger spin have less connected filaments.

A scenario compatible with these results is that where halo accretion from fewer preferred filaments/directions boosts the angular momentum increase while when several filaments contribute mass with different angular momentum tend to cancel each other out.
We also explored the assembly history of these haloes and we find that major mergers produce large changes in the halo spin, { disrupting} the spin-filament relation. The scenario of filaments regulating the spin growth of haloes is only valid in the early stages of halo formation before major mergers begin shaping the fate of halo spins.

We also study the alignment of the halo angular momentum and its connected filaments.  We found an important alignment for higher spin and haloes with few filaments. This is in good agreement with a scenario where the main contributor to the growth of the angular momentum is a single, main filament where the angular momentum aligns perpendicular to the filament, and additional filaments misaligned with the main filament tend to cancel angular momentum contributions.
{ In addition, the alignment does not depend on halo mass.}

{
The scenario proposed by \citet{2015MNRAS.452.3369C} can not be tested in this study since our analysis is centered on smaller scales, on filaments connected to halos rather than on the nearest large scale filament. In addition, we cannot resolve the small halos accreted from filaments in our simulations.
However, due to the hierarchical nature of structure formation in our Universe we expect that the ``Constrained Tidal Torque Theory'' should be valid at smaller scales. 
In this context, our halos with few filaments resemble the behavior of DM haloes flowing from saddle points ending up inside nodes of the cosmic web; they have high spin and are more perpendicular to the filament. On the other extreme, small spin halos with several filaments tend to have a more spherically symetric shape and therefore the coupling between their inertia moment and the tidal field is smaller compared to asymetric shapes, which can result in low halo angular momentum.
 To make direct comparisons of our results with the constrained TTT and other large scale filament alignment scenarios, more studies on connectivity should be done to make a direct link between the cosmic web filaments and the filaments attached to halos at virial radii scales. 

}

A natural step in the direction of the understanding of the spin-filament relation and alignments is to include baryons.  The spin alignment between the gas and DM for a few individual galaxy simulations at high redshifts was done by \citep{2015MNRAS.452..784P}. Their study is focused on objects where major mergers have not disrupted the spin of haloes.  They found that spins are well aligned at these initial stages \citep{2015MNRAS.452..784P}, therefore we expect a similar spin-filament relation for the baryons. However, huge baryonic simulations of cosmological volumes are required to investigate this, rather than individual realizations. That is work in progress to be published in an upcoming paper.

\section*{Acknowledgments}
REG was supported by Proyecto Financiamiento Basal PFB-06 'Centro de
Astronomia y Tecnologias Afines' and Proyecto Comit\'e Mixto ESO
3312-013-82.
NP acknowledges support from Fondecyt Regular 1150300.
The Geryon cluster at the Centro de Astro-Ingenieria UC was
extensively used for the calculations performed in this paper. The
Anillo ACT-86, FONDEQUIP AIC-57, and QUIMAL 130008 provided funding
for several improvements to the Geryon cluster.

%%%%%%%%%%%%%%%%%%%% REFERENCES %%%%%%%%%%%%%%%%%%
% The best way to enter references is to use BibTeX:

\bibliographystyle{mnras}
\bibliography{satbib} % if your bibtex file is called example.bib

%\input{spinfil.bbl}

% Don't change these lines
\bsp    % typesetting comment
\label{lastpage}
\end{document}